\documentclass[twocolumn,showpacs,preprintnumbers,amsmath,amssymb]{revtex4-1}
\usepackage{natbib}
\usepackage{graphicx}
\usepackage{dcolumn}
\usepackage[english]{babel}
\usepackage{amsmath}
\usepackage{amssymb}
\usepackage{tabularx}
\usepackage{multirow}
\usepackage[all]{xy}
\usepackage{appendix}
\usepackage{xspace}
\usepackage{color}
\usepackage[utf8]{inputenc}
\usepackage{titlesec}

\begin{document}

\title{Multimodal unidirectionnal pulse propagation equation}

\author{P. B\'ejot}\email{pierre.bejot@u-bourgogne.fr}
\address{Laboratoire Interdisciplinaire Carnot de Bourgogne, UMR 6303 CNRS/Université Bourgogne Franche-Comt\'e, 21078 Dijon, France}

\begin{abstract}
In this paper, the unidirectional pulse propagation equation generalized to structured media is derived. A fast modal transform linking the spatio-temporal representation of the field and its modal distribution is presented. This transform is used for solving the propagation equation by using a split-step algorithm. As an example, we present, to the best of our knowledge, the first numerical evidence of the generation of conical waves in highly multimodes waveguides.
\end{abstract}

\maketitle

\section{Introduction}

Recent progresses in optical fibers technology have renewed strong attention on the possibility of using highly multimodal waveguides to overcome  limitations inherently linked to monomode propagation \cite{1,2,3,4,5}. In parallel, gas-filled hollow waveguides are also of particular interest for their capacity of carrying and exacerbating the nonlinear propagation of ultra-intense laser pulses without any risk of material damages \cite{6,7,8}. Contrary to free-space propagation, the use of external waveguides allows to design and tailor the propagation medium properties (in particular, the frequency dispersion properties) almost at will, increasing in turn the variety and richness of the propagation dynamics to explore. In a more fundamental point of view, in analogy with the mesoscopic branch of condensed matter physics, massively multimodal propagation can be viewed as the transition between a purely quantized system in which only one or a few states (i.e., modes) can be excited and the purely continuous macroscopic physics in which the eigenstates are dense in the energy space. Pushing forward this analogy, the study of multimode waveguide could then be of great interest to study, in tabletop experiments, the physics of mesoscopic systems as optical black holes can be potentially useful for studying the properties of black holes \cite{9}.
In this context, it is then of prime importance to develop models able to capture as close as possible the propagation dynamics in such waveguiding structures in which potentially hundreds of optical modes can be excited. As already mentioned in \cite{10}, the nonlinear propagation in massively multimodal optical waveguides are characterized by several properties that are difficult to handle numerically at the same time. First, the waveguide structure is characterized by a strong refractive index contrast and can even potentially present a discontinuity in its spatial distribution (for instance, in the case of step-index or gas-filled hollow core fibers). Second, most of the time, the propagation is inherently unidirectional and over long distances, which can be very time consuming in the context of numerical simulations. Finally, the electric field propagating within the guiding structure can experience a strong reshaping in the time, spectral, and spatial domains. Up to now, most of the derived numerical models are not suited for rendering all these characteristics at the same time. Basically, there are two different approaches to tackle such guided propagation problems \cite{4}. The first one consists to fully work in a modal representation of the electric field \cite{11}. This method has, however, two main drawbacks. First, using a modal representation requires that the modes are well known. While the ones of some "simple" architectures are well known, it can become quite difficult to accurately determine the transversal shape of the modes together with their associated propagation constants in the case of complex waveguiding structures. In fact, such propagation method is very efficient in the linear regime. However, in the case of nonlinear propagation during which there are many intermodal nonlinear couplings, such kind of calculation needs to evaluate, prior to any propagation calculations, overlap integrals between any modes combinations. While such preliminary calculations are easily done in the case of weakly multimodes waveguide, it becomes very demanding both in terms of computing time and required memory if hundreds of modes are at play. For instance, in the case of a nonlinear propagation in which only a third order nonlinearity is at play (Kerr effect), the number of overlap integrals to calculate is equal to the binomial coefficient ${N+4-1}\choose{4}$, where $N$ is the number of guided modes, and then scales approximatively as the fourth power of the number of modes at play. The second kind of calculations is based on a direct representation in the space-time domain of the electric field. Basically, it consists to solve a propagation equation originally derived in the case of propagation in homogeneous media (such as the generalized nonlinear Schr{\"o}dinger equation or other similar propagation equation), in which the spatial structure of the waveguide is included \textit{a posteriori} as a correction to the source term in the Maxwell equations. This method leads to a Gross–Pitaevskii equation, in which the refractive index spatial structure acts as a potential term. As already noted in \cite{10}, this artifice nevertheless automatically adds strong approximations in the propagation equation as compared to the exact case. In this paper, we propose an alternative method that combines all advantages of the two kinds of methods. It consists to solve the exact generalized unidirectional pulse propagation equation (gen-UPPE) written in the modal basis by a split-step method in which each terms composing the propagation equation (linear propagator, nonlinear polarization, and free current in the case of propagation in a ionized medium) are calculated in the representation in which there are the most efficiently evaluated. For this, a fast modal numerical transform (in analogy with the fast Fourier or the fast Hankel transforms) is encoded, with which one can efficiently switch between the spatio-temporal and the full modal representation of the electric field for any waveguide geometry. We restrict our studies to waveguides presenting a cylindrical symmetry around the propagation axis,although our formalism can be straightforwardly applied to more complex geometries. In the first section, the derivation from the Maxwell equations of the gen-UPPE is recalled in the scalar approximation. In the second section, the fast modal transform is presented after a description of the numerical method used to evaluate the propagation modes. Finally, as an example, it is shown that conical waves can be generated in multimodal waveguides, as it is the case in bulk materials.

\section{Derivation of the Generalized Unidirectionnal Pulse Propagation Equation (Gen-UPPE)}
\subsection{Generalities and hypothesis}
Let us start from the well known wave equation of the electric field:
\begin{equation}
\overrightarrow{\triangle}\overrightarrow{E}-\frac{1}{c^2}\partial_t^2\overrightarrow{E}-\overrightarrow{\nabla}(\overrightarrow{\nabla}\cdot\overrightarrow{E})=\mu_0(\partial_t\overrightarrow{J}+\partial_t^2\overrightarrow{P}),
\end{equation}
where $\overrightarrow{P}$ and $\overrightarrow{J}$ are the medium polarization and free-charges induced current, respectively. In the picosecond and femtosecond regimes, even if atoms and molecules are ionized, freed electrons do not have the time to move far from their parent ions. Consequently, the medium remains neutral even locally so that the global charge density $\rho_T=0$ and $\overrightarrow{\nabla}\cdot\overrightarrow{D}=0$.
Accordingly, one has:
\begin{equation}
\overrightarrow{\nabla}\cdot\overrightarrow{E}=\frac{\overrightarrow{\nabla}\cdot\overrightarrow{P}}{\epsilon}.
\end{equation}
For weak nonlinearities, one has in good approximation $\overrightarrow{P}=\epsilon_0\epsilon_r\overrightarrow{E}$ so that
\begin{equation}
\overrightarrow{\nabla}\cdot\overrightarrow{E}=-\frac{\overrightarrow{\nabla}(\epsilon_r)\cdot\overrightarrow{E}}{\epsilon_r}.
\end{equation}
As long as the variations of $\epsilon_r$ remain slower than the optical wavelength, the contribution of  $\overrightarrow{\nabla}(\overrightarrow{\nabla}\cdot\overrightarrow{E})$ remains far smaller than the one of $\overrightarrow{\triangle}\overrightarrow{E}$ so that the propagation equation for the real electric field can be simplified in good approximation:
\begin{equation}
\overrightarrow{\triangle}\overrightarrow{E}-\frac{1}{c^2}\partial_t^2\overrightarrow{E}=\mu_0(\partial_t\overrightarrow{J}+\partial_t^2\overrightarrow{P}).
\end{equation}
We will also neglect the longitudinal polarization component of the electric field. Moreover, we limit our calculations to electric fields linearly polarized along a transverse direction $\overrightarrow{e}_s$: $\overrightarrow{E}=E\overrightarrow{e}_s$ but the generalization to an arbitrary transverse polarization is straightforward. The scalar component of the electric field $E$ then satisfies:
\begin{equation}
\triangle E-\frac{1}{c^2}\partial_t^2E=\mu_0\left(\partial_tJ+\partial_t^2P\right).
\label{EqOnde}
\end{equation}
For the following, it will be useful to define the Fourier Transform $\widetilde{f}(\omega)$ of a temporal function $f(t)$ and the two-dimensional Fourier Transform $\widehat{f}(\omega,k_z)$ of a function $f(t,z)$ that depends on both time and longitudinal direction $z$ as:
\begin{eqnarray}
\begin{aligned}
&\widetilde{f}(\omega)=\int{f(t)e^{i\omega t}dt}&\\
&\widehat{f}(\omega,k_z)=\iint{f(t,z)e^{i\omega t}e^{-ik_zz}dtdz}.&
\end{aligned}
\end{eqnarray}
Using a Fourier transform in both $t$ and $z$, Eq.(\ref{EqOnde}) becomes:
\begin{equation}
\left(\triangle_{\bot}+\frac{\omega^2}{c^2}-k_z^2\right)\widehat{E}=\mu_0\left(-i\omega\widehat{J}-\omega^2\widehat{P}\right),
\label{PropagationEq1}
\end{equation}
where $\triangle_{\bot}=\partial^2_r+\frac{1}{r}\partial_r$ is the transverse Laplacian when cylindrical symmetry around the propagation axis $z$ is assumed.
\subsection{Polarization}
The polarization $P$ can be written as the sum of a linear and a nonlinear contribution: $P=P_\textrm{L}+P_\textrm{NL}$ where $P_\textrm{L}$, and $P_\textrm{NL}$ accounts for the linear and nonlinear part, respectively.
\subsubsection{Linear contribution}
The calculation of $P_\textrm{L}$ is easier in the frequency domain:
$\widetilde{P_L}(r,\omega)=\epsilon_0\chi^{(1)}(r,\omega)\widetilde{E}(r,\omega)$, where $\chi^{(1)}(r,\omega)$ is linked with the linear refractive index $n$ of the material by the relation: $n(r,\omega)=\sqrt{1+\chi^{(1)}(r,\omega)}$ (assuming a cylindrical symmetry with respect to the propagation axis).
In the linear case ($J$=0 and $P_{\textrm{NL}}$=0), the propagation equation then becomes:
\begin{equation}
(\triangle_{\bot}+\frac{n^2(r,\omega)\omega^2}{c^2}+\partial^2_z)\widetilde{E}=0
\label{EqPropLin}
\end{equation}
\subsubsection{Nonlinear contribution}
Even if the functional form of the nonlinear polarization does not matter for the following derivation, one can think to the instantaneous Kerr effect:
$P_{\textrm{NL}}(r,t)=\epsilon_0\chi^{(3)}E^{3}(r,t)$ where $\chi^{(3)}$ is the third order susceptibility of the medium. In the case of molecular gases, one can have also a contribution coming from the rotational Raman-induced nonlinear polarization: $P_{\textrm{NL}}(r,t)=\rho_{\textrm{mol}}\left(\bar{\alpha}+\Delta\alpha\left[<\textrm{cos}^2\theta>-1/3\right]\right)$ (for the case of symmetric top molecules such as nitrogen or oxygen), where $\rho_{\textrm{mol}}$ is the molecular density, $\bar{\alpha}=(2\alpha_\perp+\alpha_\parallel)/3$, $\Delta\alpha=\alpha_\parallel-\alpha_\perp$ with $\alpha_\parallel$ (resp. $\alpha_\perp$) the first order polarizability of the molecule along (resp. perpendicular to) its symmetry axis, and $\theta$ is the angle between the laser polarization axis and the symmetry axis of the molecule, which nonlinearly depends on the electric field amplitude. In the case of glasses or crystals, a vibrational Raman-induced nonlinear polarization may also contribute to the materials optical response.
\subsection{Free charges induced current}
When atoms or molecules are ionized, the freed electrons (mass $m_\textbf{e}$ and electric charge $e$) moves with a velocity $\textbf{v}_\textbf{e}$ and induce a current $\textbf{J}=e\rho\textbf{v}_e$, where $\rho$ is the electrons density. Using the continuity equation and the fundamental principle of the dynamics, the induced current $J$ follows the equation of motion:
\begin{equation}
\partial_t\textbf{J}+\nu_\textbf{e}\textbf{J}=\frac{e^2\rho}{m_\textbf{e}}\textbf{E}+\Pi,
\end{equation}
with
\begin{equation}
\Pi=\frac{e}{m_\textbf{e}c}\textbf{J}\wedge\textbf{B}-\left[(\nabla.\textbf{J})\frac{\textbf{J}}{e\rho}+(\textbf{J}.\nabla)\textbf{v}_\textbf{e}\right]
\end{equation}
representing the ponderomotive forces. For intensities lower than $10^{15}$\,W/cm$^{2}$, $\Pi$ can be safely neglected.
Consequently, in the frequency space, J verifies:
\begin{equation}
\widetilde{\textbf{J}}=\frac{e^2(\nu_\textbf{e}+i\omega)}{m_\textbf{e}(\nu_e^2+\omega^2)}\widetilde{\rho\textbf{E}}.
\end{equation}
Moreover, $\rho$ is evaluated as:
\begin{equation}
\partial_t{\rho}=W(I)\left(\rho_{\textrm{at}}-\rho\right)+\frac{\sigma}{U_i}I-f(\rho),
\end{equation}
where $W(I)$ accounts for the ionization probability, $\rho_{\textrm{at}}$ is the density of neutrals, $I$ is the electric field envelope intensity, $U_i$ is the ionization potential, $f$ is a recombination function, and $\sigma$ is the inverse Bremsstrahlung cross-section defined as:
\begin{equation}
\sigma=\frac{e^2\tau_\textrm{c}}{\epsilon_0m_\textrm{e}c\left(1+\tau_\textrm{c}\omega^2_0\right)},
\end{equation}
where $\tau_\textrm{c}$ is the collision time between an electron and an neutral atom, and $\omega_0$ is the carrier frequency of the driving electric field.
\subsection{Derivation of the unidirectional propagation equation in a waveguide}
Let us recall the propagation equation driving the electric field evolution expressed in the space $(r,\omega,z)$:
\begin{equation}
\left[\triangle_{\bot}+\frac{n^2(r,\omega)\omega^2}{c^2}+\partial^2_z\right]\widetilde{E}=-i\mu_0\omega\widetilde{J}-\frac{\omega^2}{\epsilon_0c^2}\widetilde{P_{\textrm{NL}}}.
\label{PropagationEq3}
\end{equation}
Let us now assume that one can form a basis set composed of solutions $\widetilde{\varepsilon}_{\small{\mathcal{M}}}(r,\omega,z)$ of the linear propagation equation such that:
\begin{equation}
\widetilde{\varepsilon}_{\small{\mathcal{M}}}(r,\omega,z)=\widetilde{\mathcal{A}}[r,\omega,K_z(\omega)]e^{iK_z(\omega)z}.
\end{equation}
The basis set is chosen so that the basis vectors are orthogonal for the following scalar product:
\begin{equation}
\int\widetilde{\mathcal{A}}[r,\omega,K_{z_{1}}(\omega)]\widetilde{\mathcal{A}}[r,\omega,K_{z_{2}}(\omega)]rdr=\frac{\delta(K_{z_{1}}-K_{z_{2}})}{K_{z_{1}}}.
\end{equation}

Accordingly, the electric field distribution can be written as:
\begin{equation}
E(r,\omega,z)=\int\overline{E}(\omega,K_z,z)\widetilde{\mathcal{A}}[r,\omega,K_{z}(\omega)]e^{iK_z(\omega)z}K_zdK_z,
\end{equation}
with $\overline{E}(\omega,K_z,z)$ the coordinates of the electric field $E(r,t,z)$ in this basis:
\begin{equation}
\overline{E}(\omega,K_z,z)=\int E(\omega,r,z)\widetilde{\mathcal{A}}[r,\omega,K_{z}(\omega)]e^{-iK_zz}rdr.
\label{ModalTransform}
\end{equation}
The propagation equation then reads in the modal basis:
\begin{equation}
\left[\triangle_{\bot}+\frac{n^2(r,\omega)\omega^2}{c^2}+\partial^2_z\right]\overline{E}=-i\mu_0\omega\overline{J}-\frac{\omega^2}{\epsilon_0c^2}\overline{P}_{\textrm{NL}}
\label{PropagationEq4}
\end{equation}
Making a Fourier transform in $z$, the equation reduces to:
\begin{equation}
\left[K^2_z(\omega)-k_z^2\right]\widehat{\overline{E}}=-i\mu_0\omega\widehat{\overline{J}\, }-\frac{\omega^2}{\epsilon_0c^2}\widehat{\overline{P}}_{\textrm{NL}}.
\label{PropagationEq5}
\end{equation}

As a consequence, $\widehat{\overline{E}}$ verifies:
\begin{equation}
\widehat{\overline{E}}=\frac{\widehat{\overline{F}}_{\textrm{NL}}}{k_z^2-K^2_z},
\end{equation}
with
\begin{equation}
\overline{F}_{\textrm{NL}}=i\mu_0\omega\overline{J}+\frac{\omega^2}{\epsilon_0c^2}\overline{P}_{\textrm{NL}},
\end{equation}
expressed in the modal basis.
Moreover, one has:
\begin{equation}
\frac{1}{k_z^2-K^2_z}=\frac{1}{2K_z}\left(\frac{1}{k_z-K_z}-\frac{1}{k_z+K_z}\right).
\end{equation}

From now on, one can decompose the electric field as the sum of a backward and a forward propagating field which will be denoted as $\overline{E}_-$ and $\overline{E}_+$, respectively. Accordingly, neglecting the backward component, one obtains the propagation equation of $\overline{E}_+$ by using a z inverse Fourier transform. One finally obtains the unidirectional pulse propagation equation expressed in the modal basis:
\begin{equation}
\begin{aligned}
&\partial_z\overline{E}_+=iK_z\overline{E}_+&\\
&+\frac{1}{2K_z}\left(\frac{i\omega^2}{\epsilon_0c^2}\overline{P}_{\textrm{NL}}-\mu_0\omega\overline{J}\right).&
\end{aligned}
\end{equation}
Note that this equation is a generalization of the UPPE derived in bulk media. In the latter case, the Bessel functions $J_0(\sqrt{k^2(\omega)-K^2_z}r)$ form the modal basis with propagation constant $K_z$. However, by a simple change of variables, one can equivalently use the Bessel functions $J_0(k_\perp r)$ with propagation constant $K_z=\sqrt{k^2(\omega)-k^2_\perp}$. Doing this, the projection in the modal basis (Eq.\ref{ModalTransform}) reduces to the well-known Hankel-transform which can be numerically implemented very efficiently \cite{11}. Moreover, in this case, the transversal distribution of the modes is independent from $\omega$, which highly simplifies the calculations.

For numerical implementation, it is convenient to make a change of variables in order to keep the pulse centered around $t=0$ all along the propagation.
This is made by defining a sliding time origin such that the time origin corresponds for any $z$ to the time at which the pulse would be maximal if it propagated with a velocity $v$. Accordingly, one defines the new set of variables

\begin{equation}
\left\{
    \begin{array}{ll}
    \zeta=z,\\
    \tau=t-z/v.
    \end{array}
\right.
\end{equation}
The velocity $v$ can be chosen arbitrarily but a convenient choice is to use $v=v_{g_{0}}=1/\partial_\omega K_{z_{0}}$, where $K_{z_{0}}$ is the propagation constant of the fundamental guided mode. Accordingly, the partial derivative becomes:
\begin{equation}
\left\{
    \begin{array}{ll}
    \partial z=\partial\zeta-\frac{1}{v_{g_{0}}}\partial\tau,\\
    \partial\tau=\partial t.
    \end{array}
\right.
\end{equation}
In the frequency domain, this leads to:
\begin{equation}
\partial z=\partial\zeta +\frac{i\omega}{v_{g_{0}}}.
\end{equation}

Finally, the propagation equation becomes:
\begin{equation}
\partial_z\overline{E}_+=i(K_z-\frac{\omega}{v_{g_{0}}})\overline{E}_++\frac{1}{2K_z}\left(\frac{i\omega^2}{\epsilon_0c^2}\overline{P}_{\textrm{NL}}-\mu_0\omega\overline{J}\right).
\end{equation}
By replacing in the above equation, the expression of $P_{NL}$ and $J$ derived above, one finally obtains:
\begin{equation}
\begin{aligned}
&\partial_z\overline{E}_+=i(K_z-\frac{\omega}{v_{g_{0}}})\overline{E}_+&\\
&+\frac{1}{2K_z}\left(\frac{i\omega^2}{c^2}\overline{\chi^{(3)}E_+^3}-\omega\frac{e^2}{\epsilon_0m_ec^2}\frac{\nu_e+i\omega}{\nu_e^2+\omega^2}\overline{\rho E_+}\right).&
\end{aligned}
\label{EqProp}
\end{equation}

\subsubsection{Complex representation of the UPPE}

It is often simpler to use a complex representation of the electric field. Let us define $\varepsilon$ such that the real electric field writes
\begin{equation}
E=\frac{\varepsilon+\varepsilon^*}{2},
\end{equation}
making the change of variable $\xi=\sqrt{\frac{\epsilon_0cn}{2}}\varepsilon$ such that $I(t)=|\xi|^2(t)$ represents the pulse intensity (i.e. the average over a few optical cycles of the Pointing vector), one obtains the following equation for the complex electric field:

\begin{equation}
\begin{aligned}
&\partial_z\bar{\xi}=i(K_z-\frac{\omega}{v_{g_{0}}})\bar{\xi}&\\
&+\frac{1}{K_z}\left[\frac{i\omega^2}{c^2}\left(\overline{n_2|\xi|^2\xi}+\overline{n_2\frac{\xi^3}{3}}\right)-\omega\frac{e^2}{2\epsilon_0m_ec^2}\frac{\nu_e+i\omega}{\nu_e^2+\omega^2}\overline{\rho\xi}\right],&
\end{aligned}
\end{equation}
where $n_2=\frac{3}{4}\frac{\chi^{(3)}}{\epsilon_0cn}$ is the nonlinear refractive index of the medium. Moreover, taking into account the optical losses induced by ionization, one finally has:
\begin{equation}
\begin{aligned}
&\partial_z\bar{\xi}=i(K_z-\frac{\omega}{v_{g_{0}}})\bar{\xi}-\rho_{\textrm{at}}U_i\overline{\frac{W_I\xi}{2I}}&\\
&+\frac{1}{K_z}\left[\frac{i\omega^2}{c^2}\left(\overline{n_2|\xi|^2\xi}+\overline{n_2\frac{\xi^3}{3}}\right)-\omega\frac{e^2}{2\epsilon_0m_ec^2}\frac{\nu_e+i\omega}{\nu_e^2+\omega^2}\overline{\rho\xi}\right].&
\end{aligned}
\label{EqPropComplex}
\end{equation}

Up to now, we have supposed the existence of a modal basis without explicitly exhibiting its mathematical form. Such a basis and the related propagation constants $K_z$ are well known for simple waveguide architectures (e.g. step-index fibers or free space propagation). The next section will describe how the modal basis can be calculated in a more general case and how one can switch from a description of the field in the space $(r,t)$ to its modal expansion. This change of representation will be particularly useful for numerically solving the propagation equation.
\subsection{Modal expansion}
In this subsection, we will discuss how the transversal modes can be calculated in the linear propagation case. More particularly, one wishes to find the transverse distributions of the electric field satisfying:
\begin{equation}
\left(\triangle_\bot+\frac{n^2(r,\omega)\omega^2}{c^2}\right)\widehat{\varepsilon}(r,k_z,\omega)=k_z^2\widehat{\varepsilon}(r,k_z,\omega).
\end{equation}
Finding the propagation modes then amounts to search the eigenvectors of the operator $\Box_\bot\equiv\triangle_\bot+\frac{n^2(r,\omega)\omega^2}{c^2}$ and the associated eigenvalues corresponding to the square of the propagation constants of these modes. To this aim, we choose to express the operator $\Box_\bot$ in a basis composed of the zeroth-order Bessel functions $J_0(k_\bot r)$. Note that this basis is orthogonal for the following scalar product:
\begin{equation}
\int rJ_0(k_{\bot_{1}}r)J_0(k_{\bot_{2}}r)dr=\frac{\delta(k_{\bot_{1}}-k_{\bot_{2}})}{k_{\bot_{1}}}.
\end{equation}
In this basis, one can express any radial function $f(r)$ as:
\begin{equation}
f(r)=\int f(k_{\bot})J_0(k_{\bot}r)k_{\bot} dk_{\bot},
\end{equation}
which corresponds in fact to the Hankel transform (denoted hereafter as TH) and with
\begin{equation}
f(k_{\bot})=\int f(r)J_0(k_{\bot}r)rdr.
\end{equation}
In particular, one can write $\Box_\bot[J_0(k_{\bot}r)]$ as
\begin{equation}
\Box_\bot[J_0(k_{\bot}r)]=\int I(k'_{\bot},k_{\bot})J_0(k'_{\bot}r)k'_{\bot}dk'_{\bot},
\end{equation}

where
\begin{eqnarray}
\begin{aligned}
I(k'_{\bot},k_{\bot})&=\int rJ_0(k_{\bot}r)\triangle_\bot J_0(k'_{\bot}r)dr&\\ \nonumber
&+\frac{\omega^2}{c^2}\int n^2(r,\omega)J_0(k_{\bot}r)J_0(k'_{\bot}r)rdr&\\ \nonumber
\, &=-k_{\bot}^2\frac{\delta(k_{\bot}-k'_{\bot})}{k_{\bot}}+\frac{\omega^2}{c^2}I_1(k_{\bot},k'_{\bot}),&\nonumber
\end{aligned}
\end{eqnarray}
and where $I_1(k'_{\bot},k_{\bot})=\int n^2(r,\omega)J_0(k_{\bot}r)J_0(k'_{\bot}r)rdr$.

In this basis, $\Box_\bot(f)$ writes as:
\begin{eqnarray}
\begin{aligned}
&\Box_\bot[f](r)=\int f(k_{\bot_{1}})\Box_\bot\left[J_0(k_{\bot_{1}}r)\right]k_{\bot_{1}} dk_{\bot_{1}}&\\
&=\iint f(k_{\bot_{1}})I(k_{\bot_{1}},k_{\bot_{2}})J_0(k_{\bot_{2}}r)k_{\bot_{2}}k_{\bot_{1}}dk_{\bot_{2}}dk_{\bot_{1}}&\\
&=\int \mathcal{F}(k_{\bot_{2}})J_0(k_{\bot_{2}}r)k_{\bot_{2}}dk_{\bot_{2}},&
\end{aligned}
\end{eqnarray}
where $\mathcal{F}(k_{\bot_{2}})=\int f(k_{\bot_{1}})I(k_{\bot_{1}},k_{\bot_{2}})k_{\bot_{1}}dk_{\bot_{1}}$ corresponds to the Hankel transform of $\Box_\bot(f)$:
\begin{equation}
TH[\Box_\bot(f)](k_{\bot_{2}})=\mathcal{F}(k_{\bot_{2}}).
\end{equation}

We look for functions $\varepsilon$ such that $\Box_\bot(\varepsilon)=k^2_z\varepsilon$. Expressed in the Hankel space, it reads

\begin{equation}
\boxed{
k^2_z\varepsilon(k_{\bot_{2}})=\int\varepsilon(k_{\bot_{1}})I(k_{\bot_{1}},k_{\bot_{2}})k_{\bot_{1}}dk_{\bot_{1}},
}
\label{Kernel0}
\end{equation}
which corresponds to an homogeneous Fredholm integral equation of the second kind with the real kernel $K\left(k_{\bot_{1}},k_{\bot_{2}}\right)=k_{\bot_{1}}I\left(k_{\bot_{1}},k_{\bot_{2}}\right)$.
Finding the eigenmodes of the linear propagation equation then amounts to look for the eigenvalues and eigenvectors of the kernel $K$.
\subsubsection{Basic examples}
Let us consider the bulk case, i.e. $n(r)=n$. In this case, the kernel $K$ takes the form
\begin{equation}
K\left(k_{\bot_{1}},k_{\bot_{2}}\right)=\left(\frac{n^2\omega^2}{c^2}-k^2_{\bot_{1}}\right)\delta(k_{\bot_{1}}-k_{\bot_{2}}),
\end{equation}
and is consequently \textit{diagonal}. One then directly obtains that the functions $\varepsilon(k_{\bot})=\frac{\delta(k_{\bot}-K_{\bot})}{k_{\bot}}$ are solutions of the equation and the eigenvalues read $k^2_z=\frac{n^2\omega^2}{c^2}-K^2_{\bot}$. In the direct space, one retrieves that Bessel functions $J_0(K_{\bot}r)$ are eigenmodes of the propagation with propagation constant $\sqrt{\frac{n^2\omega^2}{c^2}-K^2_{\bot}}$.\\
\begin{figure}[h!]
	\includegraphics[width=9cm,keepaspectratio]{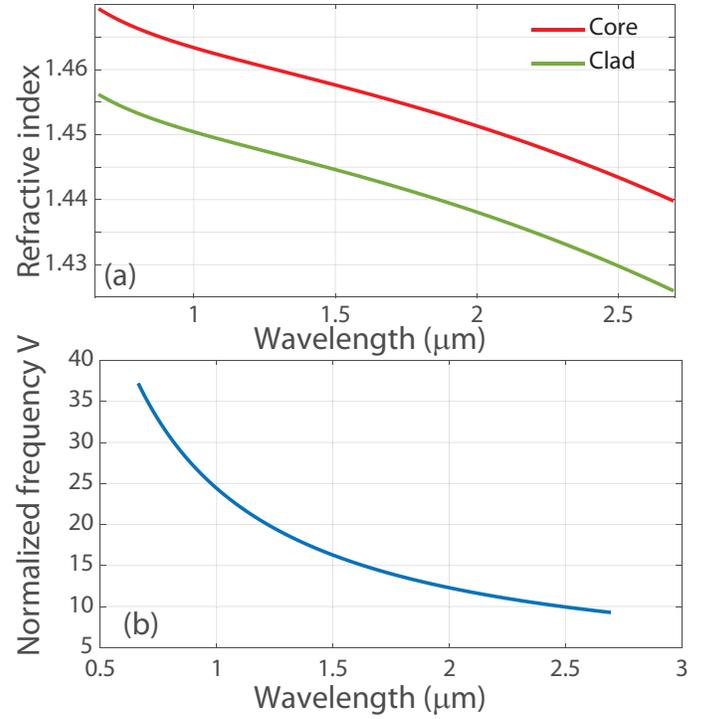}
	\caption{(a) Refractive index of the core (red) and clad (green) of the optical fiber and (b) associated normalized frequency as a function of wavelength.}
	\label{Fig1}
\end{figure}
Let us now consider the case of optical fibers. The problem is then confined to a restricted transverse space (delimited by the clad radius $R$) so that one can reasonably consider that the electric field vanishes outside the clad.
In the interval $[0,R]$, the field f(r) can be expanded as a Fourier-Bessel series (this means that the Hankel transform of $f$ is discrete):

\begin{equation}
f(r)=\sum_{j=1}^\infty {f_j}F_j(r),
\end{equation}
where $F_j(r)=\frac{\sqrt{2}}{RJ_1(\alpha_j)}J_0(\alpha_j\frac{r}{R})$, $\alpha_j$ is the $j^{\textrm{th}}$ roots of the zeroth-order Bessel function and $f_j$ is given by:
\begin{equation}
f_j=\int_0^R rf(r)F_j(r)dr.
\end{equation}
In the orthonormal basis $(F_j)$, $f$ can then be expressed as a vector called hereafter $F$.
Let us evaluate $\Box_\bot f$ in this basis:
\begin{eqnarray}
\begin{aligned}
\Box_\bot f(r)&=\sum_{j=1}^\infty f_j \Box_\bot[F_j(r)]&\\
&=\sum_{j=1}^\infty f_j (\triangle_\bot+\frac{n^2(r,\omega)\omega^2}{c^2})F_j(r).&
\end{aligned}
\label{BoxBessel}
\end{eqnarray}
The first term in the sum, corresponding to the transverse Laplacian, reads:
\begin{equation}
\triangle_\bot F_j(r)=-\frac{\alpha_j}{R^2}F_j(r).
\label{triangleBessel}
\end{equation}
The second term in the sum can be developed as a Fourier-Bessel series:
\begin{equation}
\frac{n^2(r,\omega)\omega^2}{c^2}F_j(r)=\sum_{k=1}^\infty K_{kj}F_k(r).
\label{indexBessel}
\end{equation}
Combining Eqs. (\ref{BoxBessel}-\ref{indexBessel}), one obtains:
\begin{equation}
\Box_\bot f(r)=\sum_{k=1}^\infty\left(\sum_{j=1}^\infty \left(K_{kj}-\frac{\alpha^2_j}{R^2}\delta_{k,j}\right)f_j(r)\right)F_k(r),
\label{BoxBesselFinal}
\end{equation}
where $\delta_{k,j}$ is the Kronecker delta. Equation (\ref{BoxBesselFinal}) can be written in (infinite) matrix notation as:
\begin{equation}
\Box_\bot f=MF,
\end{equation}
where $M_{kj}=K_{kj}-\frac{\alpha^2_j}{R^2}\delta_{k,j}$ and where $F$ is the coordinate vector of $f$ expressed in the $\left(F_j\right)$ basis set.\\
Finding the modes of propagation then amounts to find the eigenvalues and associated eigenvectors of $M$. Note that $M$ being a real-valued symmetric matrix, it can be diagonalized. Moreover, the eigenvectors are orthogonal and the associated eigenvalues are all real.
\\
For instance, in the case of a step-index optical fiber where the core has a size $R_1$ and where the core (resp. clad) refractive index is $n_0(\omega)$ (resp. $n_1(\omega)$), $K_{kj}$ is given by:
\begin{eqnarray}
\begin{aligned}
&K_{k\neq j}=\mathcal{K}\frac{\alpha_jJ_1(\alpha_j\frac{R_1}{R})J_0(\alpha_k\frac{R_1}{R})-\alpha_kJ_1(\alpha_k\frac{R_1}{R})J_0(\alpha_j\frac{R_1}{R})}{\left(\alpha_j^2-\alpha_k^2\right)J_1(\alpha_j)J_1(\alpha_k)}&\\ \nonumber
&K_{jj}=\frac{\omega^2}{c^2}\left(n_1^2+(n_0^2-n_1^2)\left(\frac{R_1}{R}\right)^2\frac{J^2_0(\alpha_j\frac{R_1}{R})+J^2_1(\alpha_j\frac{R_1}{R})}{J_1^2(\alpha_j)}\right),& \nonumber
\end{aligned}
\end{eqnarray}
where $\mathcal{K}=\frac{2R_1}{R}\frac{\omega^2}{c^2}\left(n_0^2-n_1^2\right)$.

\begin{figure}[h!]
	\includegraphics[width=9cm,keepaspectratio]{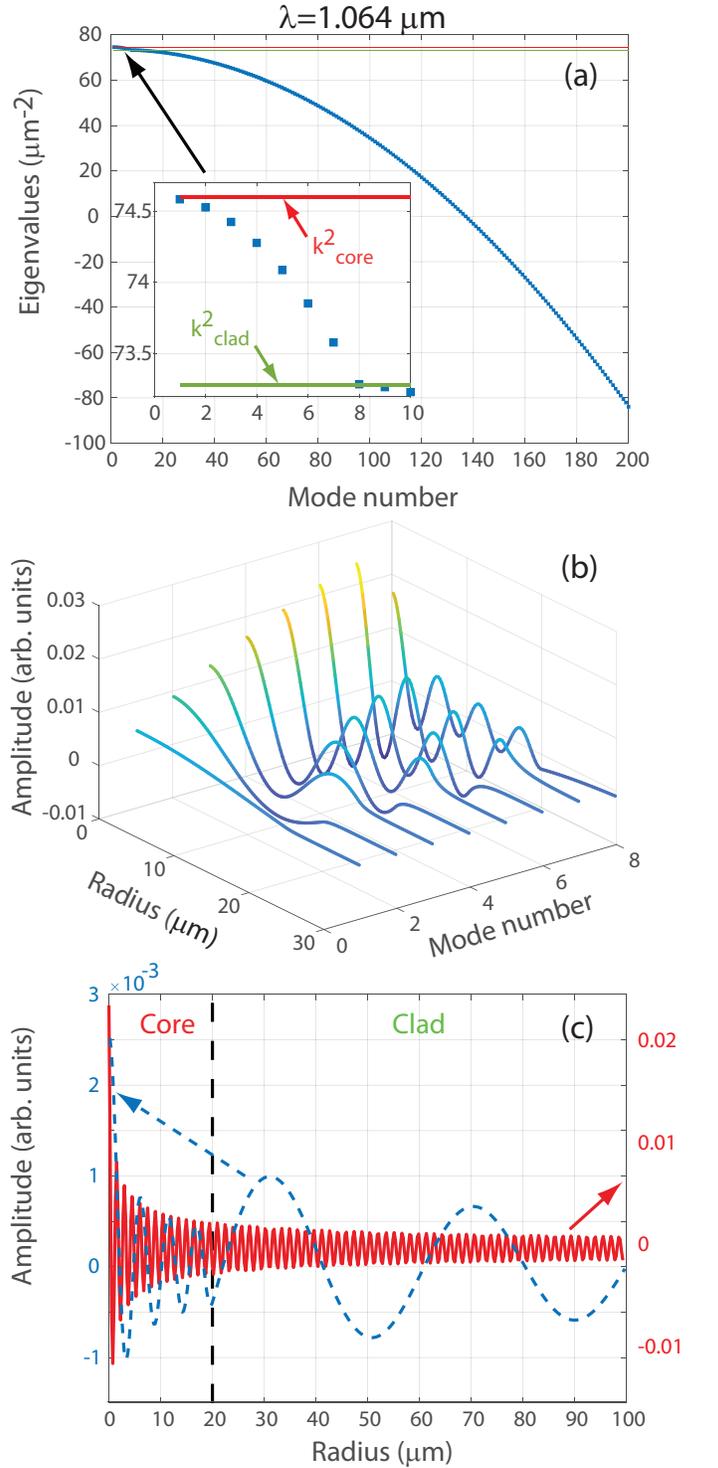}
	\caption{(a) Eigenvalues returned by the diagonalization of $M$ at $\lambda=1.064\mu$m for the step-index optical fiber. The red (resp. green) lines corresponds to the square of the wavevectors of the core (resp. clad) of the fiber. (b) Radial distributions of the eight guided modes and (c) example of a propagative clad mode (dashed blue) and of an evanescent mode (solid red).}
	\label{Fig2}
\end{figure}

In the simple case where $n_0=n_1$, $K$ (and consequently $M$) becomes diagonal and the eigenvalues $\left(k^2_j\right)$ form a countably infinite set:
\begin{equation}
k^2_j=\frac{n_0^2\omega^2}{c^2}-\frac{\alpha_j^2}{R^2}, j=1,2,...
\end{equation}
The main difference with the infinite medium case is that the possible eigenvalues are now quantified because of the finite extension of the problem.
Numerically speaking, we assume that the infinite series can be replaced with good accuracy by a finite expansion up to the N$^{\textrm{th}}$ order:
\begin{equation}
f(r)=\sum_{n=1}^N {f_n}J_0(\alpha_n\frac{r}{R}).
\end{equation}
In this case, the matrix size becomes finite and one can easily evaluate numerically the eigenvalues and eigenvectors of $M$. The diagonalization procedure will return $N$ orthogonal and normalized eigenvectors $v$ that form an orthonormal basis. In order to express the electric field in this eigenbasis, calling $V$ the matrix filled with the coordinates of the eigenvectors $v$ in the Bessel basis, an electric field whose coordinates are $F$ in the bessel basis (which corresponds to the fast Hankel transform of the electric field) will write $F'=VF$ in the eigenbasis, where $F'$ are the coordinates of the electric field in this eigenbasis. Numerically speaking, the function $f$ will be evaluated at particular values of $r$ denoted $r_j$ ($1\leq j\leq N$), which will form a vector $F_j$. In order to go from this representation of the field to the one in the eigenbasis, all one needs is to express the vector $F_j$ in the Bessel basis (i.e., to evaluate the fast Hankel transform of $F_j$) and then to apply to the new vector $\mathcal{F}$ the transformation $F'=V\mathcal{F}$. Numerically, the Hankel transform is performed by multiplying the vector $F_j$ by a unitary matrix $H$. Then, one has:

\begin{eqnarray}
\begin{aligned}
&F'=VHF_j,&\\
&F_j=H^{-1}V^{-1}F'.&
\end{aligned}
\label{ModetoR}
\end{eqnarray}
Note that, excepted particular cases such as the free space case, the change of basis matrix $V$ depends on the frequency $\omega$, which can be made in parallel for every frequency.
\subsubsection{Numerical examples}
\paragraph{Step-index optical fiber}

We will here consider a fused silica step-refractive index optical fiber with a core radius $R_1=20\,\mu$m. The refractive index of the core and the clad are shown in Fig.\ref{Fig1}(a). The associated normalized frequency $V$ is shown in Fig.\ref{Fig1}(b). Using a decomposition over the $N=200$ first Bessel functions and limiting the space to a radius $R=100\,\mu$m, the diagonalization of $M$ returns $N$ eigenvalues for each wavelength. For instance, those obtained for $\lambda=1.064\,\mu$m are displayed in Fig.\ref{Fig2}(a). The returned eigenvalues $k^2_z$ can be sorted in three distinct categories:
\begin{figure}[h!]
	\includegraphics[width=9cm,keepaspectratio]{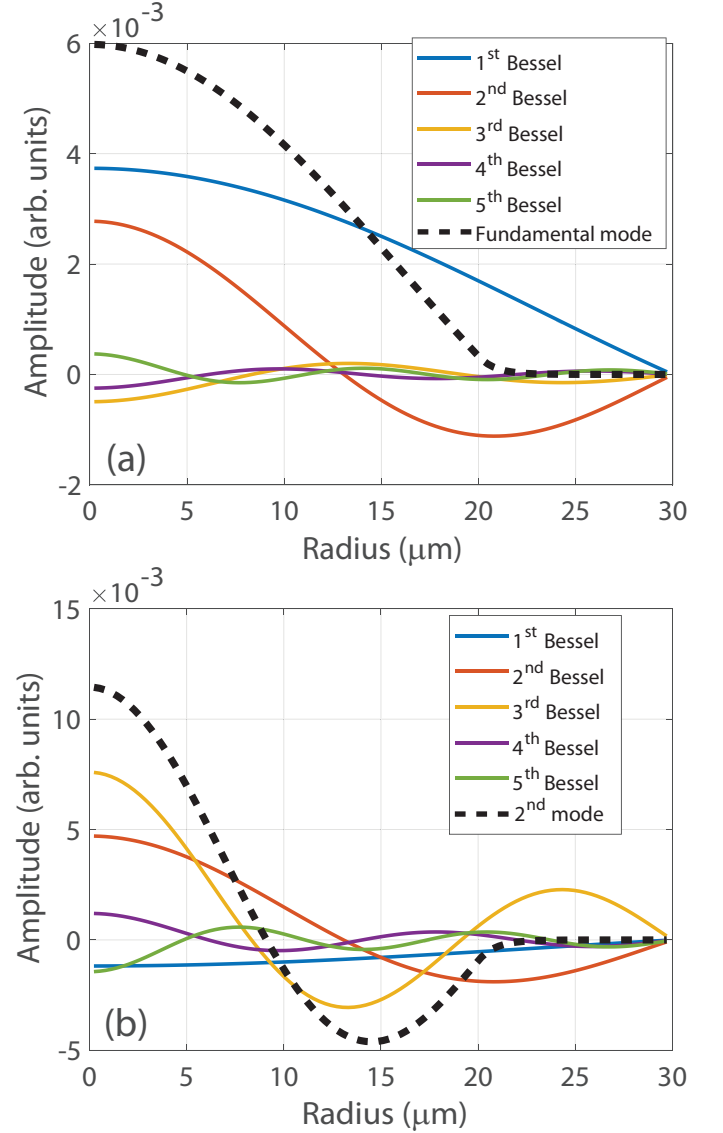}
	\caption{(a) First fifth decomposition functions of the fundamental and (b) the second mode of the step-index optical fiber over the Bessel basis.}
	\label{Fig1b}
\end{figure}
\begin{equation}
\left\{
    \begin{array}{lll}
       k^2_{\textrm{clad}}\leq k^2_z\leq k^2_{\textrm{core}},\\
       0 \leq k^2_z \leq k^2_{\textrm{clad}},\\
       k^2_z \leq 0.
    \end{array}
\right.
\end{equation}
\begin{figure}[h!]
	\includegraphics[width=9cm,keepaspectratio]{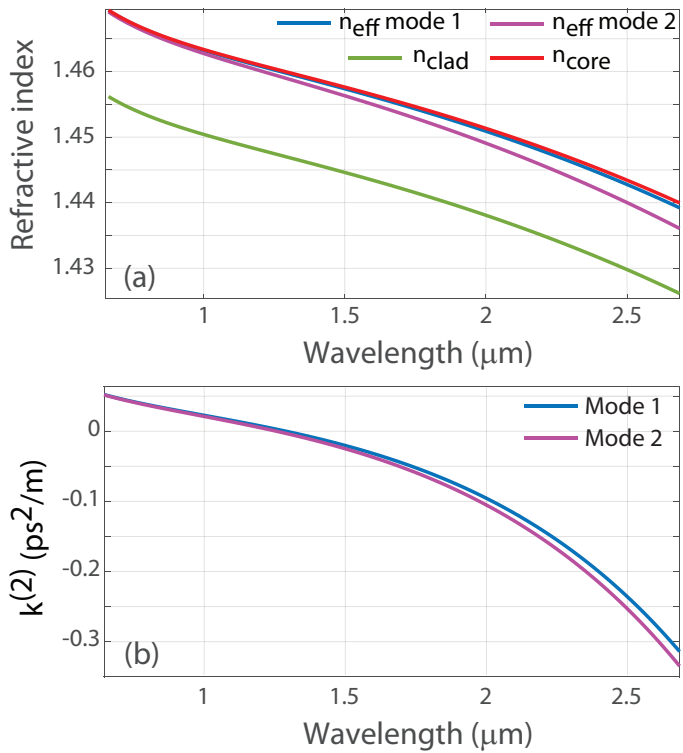}
	\caption{(a) Effective refractive index of the fundamental (blue) and second (pink) guided modes of the hollow core fiber filled with 5 bar of argon compared to the refractive index of the core (red) and clad (green). (b) Dispersion curve of the fundamental (blue) and second (pink) guided modes.}
	\label{Fig5}
\end{figure}
The first category of modes corresponds to the modes well confined within the core of the fiber (that we will call \textit{guided modes} in the following). In the present case, there are eight guided modes at this wavelength. Using Eq.\ref{ModetoR}, the spatial distributions of these modes (shown in Fig.\ref{Fig2}(b)) can be easily retrieved. The modes whose eigenvalues are in the range $[0-k^2_{\textrm{clad}}]$ are propagative modes whose distributions are not confined in the core (that we will call \textit{propagative clad modes}). A relatively important part of the energy contained in these modes propagates also in the fiber clad. The last category of modes whose eigenvalues are negative are evanescent modes. Indeed, since one has $k^2_z<0$, the associated propagation constant $k_z$ will be purely imaginary. Accordingly, those modes do not propagate but are exponentially attenuated during the propagation. Figure \ref{Fig2}(c) shows an example of such modes. The procedure of diagonalization is then performed for each wavelength giving the weight of each bessel function composing the basis for building the modes of the waveguide. For instance, Figs.\ref{Fig1b}(a,b) give the decomposition of the first two core guided modes on the used bessel basis. As shown, the waveguide modes can then be reconstructed by summing the bessel functions composing the basis weighted by the coordinate of the mode on the basis. The eigenvalue $k^2_z$ related to each eigenmode can be used so as to determine an effective refractive index $n_{\textrm{eff}_{0}}=ck_{z}(\omega)/\omega$ experienced by the mode. Such effective indices are shown in  Fig.\ref{Fig5}(a) for the two first guided modes. Moreover, one can also determine the dispersion of the fiber by evaluating for each mode the group velocity dispersion $k^{(2)}(\omega)=\partial^2_{\omega} k_z$ (see Fig.\ref{Fig5}(b)).

\begin{figure}[h!]
	\includegraphics[width=9cm,keepaspectratio]{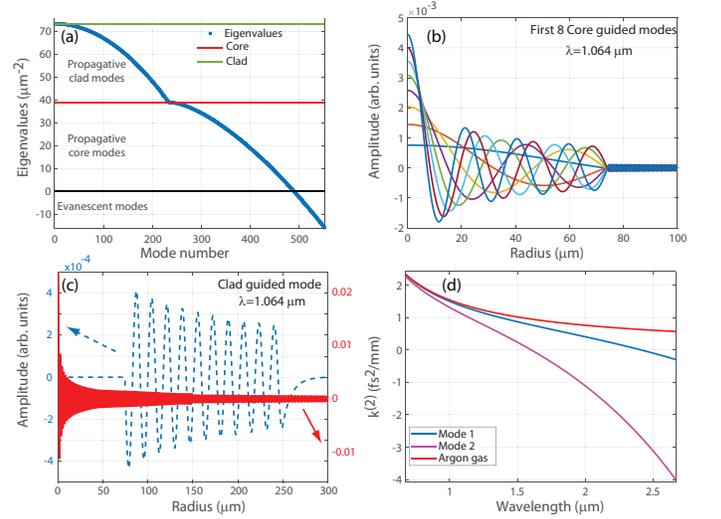}
	\caption{(a) Eigenvalues of the hollow fused silica fiber filled with 5 bar argon at $\lambda=$1.064$\mu$m. (b) Radial distribution of the first eight core guided modes. (c) Example of radial distribution of a clad guided mode (dashed blue) and an evanescent mode (red solid line) at $\lambda=1.064\,\mu m$. (d) Group velocity dispersion of the first (blue) and of the second (pink) propagation mode compared to the one of argon (red) taken at same pressure (5\,bar).}
	\label{Fig7}
\end{figure}

\paragraph{Hollow core fiber}
Let us now consider the case of a hollow core fiber. Such fiber consists of a glass clad (here we will use a fused silica cladding) in which a hole of size $R_1$ is filled with a gas (here we will use argon) at some pressure (here 5 bar). In the following, we will consider $R_1$=75$\mu$m. The situation is slightly different from a conventional optical fiber since the refractive of the core in the present case is much lower than the one of the clad. Figure \ref{Fig7}(a) shows the eigenvalues of the hollow fiber. Here again, three distinct kinds of modes can be exhibited according to their respective eigenvalues:
\begin{equation}
\left\{
    \begin{array}{lll}
       k^2_{\textrm{core}}\leq k^2_z\leq k^2_{\textrm{clad}},\\
       0 \leq k^2_z \leq k^2_{\textrm{core}},\\
       k^2_z \leq 0.
    \end{array}
\right.
\end{equation}
The first kind of modes ($k^2_{core}\leq k^2_z\leq k^2_{clad}$) is now the \textit{propagative clad modes}, whose energy is mainly located in the clad.
The second kind of modes ($0 \leq k^2_z \leq k^2_{core}$) corresponds to the \textit{propagative core modes}. However, contrary to the case of optical fibers, these mode are less confined within the core. A non-negligible part of the energy is also located in the clad so that these modes are often called \textit{leaky modes}. The third kind of modes ($k^2_z \leq 0$) corresponds to \textit{evanescent modes} as it is the case in optical fibers. Figures \ref{Fig7}(b-c) shows examples of the three kinds of optical modes. While, in the present case, there is more than 700 modes that are propagative core modes, in practice, far less modes are necessary to fairly reproduce the propagation. Figure \ref{Fig7}(d) shows the dispersion curve of the two first guided modes.
\section{Numerical solving of the Generalized Unidirectional Propagation Equation}
After having retrieved the optical modes of the waveguide, their associated propagation constants and the matrix allowing to go back and forth between the eigenbasis to the $(r,t)$ representation of the field, one has all the tools for solving the propagation equation [Eq.\ref{EqProp} or Eq.\ref{EqPropComplex}].
\subsection{Numerical strategy}
At first glance, one immediately notices that the Gen-UPPE can be put in the general form
\begin{equation}
\partial_z\bar{\xi}=\mathcal{L}_{\textrm{L}}[\bar{\xi}]+\bar{\mathcal{L}}_{\textrm{NL}}[\xi],
\end{equation}
where $\mathcal{L}_{\textrm{L}}$ (resp. $\bar{\mathcal{L}}_{\textrm{NL}}$) represents the linear (resp. nonlinear) propagation operator. Strictly speaking, those two operators can in principle be written in any representation basis (eigenbasis, Fourier-Hankel basis or in the $(r,t)$ domain). However, depending on the used basis, their evaluation can be either very simple or extremely complex and numerical resources demanding. As far as the linear operator is concerned, its evaluation is extremely simple in the eigenbasis since it amounts to simply multiply each coordinate by its propagation constant. On the contrary, the nonlinear operator explicitly depends on the electric field written in the $(r,t)$
domain. As a consequence, all terms of kind $f(r,t)$ are far simpler to evaluate with this representation. Accordingly, it becomes obvious that a split-step algorithm is well adapted for solving the generalized UPPE. Accordingly, for evaluating the field at a propagation distance $z+dz$, the numerical is divided into three distinct stages.
The first stage is a linear propagation of the field over a length $dz/2$. At this stage, the field is represented in the eigenbasis and the equation to solve is
\begin{equation}
\partial_z\bar{\xi}=iK_Z\bar{\xi},
\end{equation}
in which $K_Z$ is a diagonal matrix representing the propagation constant of each mode so that the propagated electric field writes
\begin{equation}
\bar{\xi}(z+dz)=\textrm{e}^{idz/2K_Z}\bar{\xi}(z).
\end{equation}
The second stage consists to make a nonlinear propagation over a full step $dz$. In order to do this, one simply evaluates at this stage the electric field in a space-time representation by operating the fast modal transform presented above. Knowing $\xi(r,t)$, one can easily calculate the nonlinear term (typically by calculating $|\xi(r,t)|^2\xi(r,t)$) and then use an integration scheme such as a 4$^{th}$ order Runge Kutta scheme or equivalent so as to integrate the nonlinear part of the equation.
Finally, one repeats the first stage over an half propagation step. In fact, the algorithm is similar in many ways to the well-known and widely used split-step Fourier in propagation within monomode fibers or to the split-step Hankel-Fourier algorithm used in propagation problems in bulks when assuming a cylindrical symmetry. Indeed, in these two cases, the functions $\{e^{i\omega t}\}_\omega$ and $\{e^{i\omega t}J_0(k_\bot r)\}_{\omega,k_\bot}$ coincide with the eigenfunctions of the propagation medium.
Note that one could be tempted to solve the linear propagation term directly in the Fourier-Hankel space (stage 1 and 3 described above). Indeed, in this case, the field is represented in the Fourier-Hankel space (i.e., the Fourier-Hankel transform of $\xi(r,t)$ denoted hereafter $\check{\xi}(k_r,\omega)$) and the equation to solve is
\begin{equation}
\partial_z\check{\xi}=i\check{K}_Z\check{\xi},
\end{equation}
where $\check{K}_Z$ is now the linear operator expressed in the Fourier-Hankel space and reads
\begin{equation}
\check{K}_Z=VK_ZV^{-1},
\end{equation}
where $V$ is the transformation matrix between the modal and the Hankel basis (i.e., the matrix filled with the coordinates of the modes in the Bessel basis). As a consequence, one has:
\begin{equation}
\check{\xi}(z+dz)=e^{i\check{K}_Zdz}\check{\xi}(z).
\end{equation}
However, contrary to the case in which the calculation is performed in the eigenbasis, $\check{K}_Z$ is not diagonal anymore. In this case, the calculation of its exponential becomes very time consuming. As a consequence, one can approximate it by its Pad\'e approximant of order [1/1] which has the numerical advantage to be an unitary operator:
\begin{equation}
\check{\xi}(z+dz)\simeq\frac{1+i\check{K}_Zdz/2}{1-i\check{K}_Zdz/2}\check{\xi}(z).
\end{equation}
Numerically, it is more convenient to transform this equality as:
\begin{equation}
\left(1-i\check{K}_Zdz/2\right)\check{\xi}(z+dz)\simeq\left(1+i\check{K}_Zdz/2\right)\check{\xi}(z).
\end{equation}
The right-hand side of the above equation is reduced to a matrix-vector multiplication, the output result being a vector $b$. Then, one has to solve the equation system
\begin{equation}
AX=b,
\end{equation}
where $A=1-i\check{K}_Zdz/2$ and $X=\check{\xi}(z+dz)$. For this purpose, one can use the bi-conjugate gradient stabilized (BICGSTAB) or the generalized minimal residual (GMRES) methods. However, even if there are efficient algorithms, it appears that such a way to solve the linear step is numerically more expensive than the one based on the representation of the field in the eigenbasis.
\subsection{Propagation example: Fishwave formation in multimode step-index optical fibers close to the zero-dispersion wavelength}
\begin{figure}[h!]
	\includegraphics[width=9cm,keepaspectratio]{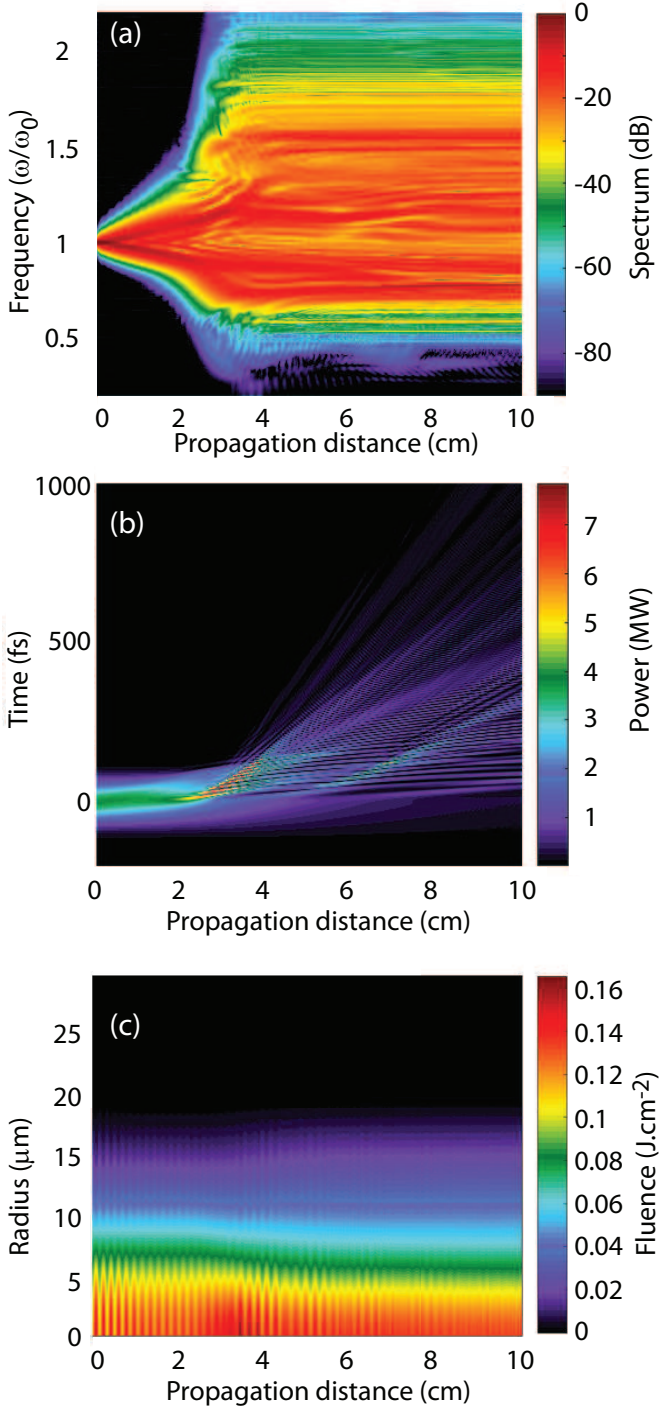}
	\caption{(a) Evolution of the integrated spectrum, (b) the instantaneous power, and (c) the fluence along the propagation in a 20\,$\mu$m step-index silica fiber.}
	\label{Fig8}
\end{figure}

Few years ago, a strong interest has been devoted to the study of conical waves formation in dispersive nonlinear bulk media \cite{13,14,15,16}. Such waves are particular solutions of the propagation equation in bulk media with the notable feature of being stationary, i.e. non-dispersive and non-diffractive, in both the linear and nonlinear regimes. Such waves are characterized by their typical hyperbolic or elliptic shapes in the angular-spectral plane depending on the dispersion curve of the bulk material. Typically, X-waves (resp. O-waves) are generated when the pump laser lies in the normal (resp. anomalous) dispersion regime. The particular situation of a pump whose central frequency lies close to the zero dispersion wavelength of the material leads to the formation of a conical waves sharing features common to both normal and anomalous dispersion regimes. This case is called a fish wave since its angular-frequency shape exhibits an elliptic core and an hyperbolic tail. To the best of our knowledge, the study of conical waves formation has been limited up to now to the case of propagation within bulk media, i.e., in media in which an infinite and dense number of modes can propagate, even if a few works have already suggest the potential existence of such waves in waveguides (see, e.g., \cite{17}). Here, we numerically show that multimode waveguides can indeed support the generation of waves whose characteristics are very similar to those of conical waves generated in bulk media. In particular, it is shown that a fish-like wave can be spontaneously formed in a step-index multimode fused silica fiber when the central frequency of the pulse lies close to the zero-dispersion frequency. Observing the formation of such waves in multimode waveguides is particularly interesting because it constitutes a typical example of the transition from the quantized physics of a few mode system (propagation within a monomode or a weakly multimode waveguide) to those of a purely continuous one (propagation through a bulk medium).\\
In this example, we considered a 1.3\,$\mu$m 100\,fs 400\,nJ laser pulse initially coupled in the fundamental core guided mode of a $R$=20\,$\mu$m step-index optical fiber. The associated peak power is about 3.7\,MW. Note that the wavelength $1.3\mu$m corresponds to the zero dispersion wavelength of this fiber. The propagation dynamics were calculated by solving Eq.(\ref{EqPropComplex}) without the ionization term ($\rho=0$) and neglecting the term responsible for third-harmonic generation. Moreover, the Kerr effect was modified so as to take into account the vibrational Raman contribution to the Kerr effect:
\begin{equation}
\begin{aligned}
&\partial_z\bar{\xi}=i(K_z-\frac{\omega}{v_{g_{0}}})\bar{\xi}&\\
&+\frac{in_2\omega^2}{c^2K_z}\left[\left(1-f_\textrm{R}\right)\overline{|\xi|^2\xi}+f_\textrm{R}\overline{\left(\int{R(\tau)|\xi|^2(t-\tau)d\tau}\right)\xi}\right],&\nonumber
\end{aligned}
\end{equation}
where $f_\textrm{R}=0.18$ is the relative contribution of the Raman effect to the nonlinearity and $R(t)$ is the Raman impulse response function \cite{18}.
Figure\,\ref{Fig8}(a) displays the evolution of the pulse spectrum (integrated over the full fiber transverse section) all along the propagation within the 10\,cm fiber. During the first three centimeters, the spectrum divides into two branches almost symmetric with respect to the pump initial frequency. After this first stage, the spectrum undergoes a fast broadening so as to form a supercontinuum spanning over about an octave.
The formation of the supercontinuum is accompanied together with a strong break up of the pulse in the temporal domain as it can be noticed in Fig.\,\ref{Fig8}(b). In particular, one can notice several temporally localized \textit{wavepackets} propagating slower than the pump pulse. In the spatial domain, the multimode nature of the propagation dynamics manifests itself if one looks at how the pulse is spatially distorted along the propagation. For instance, the fluence (i.e., the energy per surface unit) oscillates almost periodically all along the propagation. In fact, the oscillation period $L_{\textrm{osc}}$ corresponds to the \textit{coherence length} $L_{\textrm{coh}}$=$\pi/\Delta k$ with $\Delta k=k_{z_1}-k_{z_0}$, where $k_{z_0}$ (resp. $k_{z_0}$) is the propagation constant of the fundamental (second) transversal mode at the fundamental frequency. This then means that the nonlinear propagation leads to the generation of higher spatial modes (and in particular the second transversal mode). The generation of higher order transversal modes can be easily understood by noting that, if the transversal distribution of the field $\xi(r)$ corresponds initially to the fundamental mode, the nonlinear term proportional to $|\xi(r)|^2\xi(r)$ does not share the same spatial distribution than $\xi(r)$. In other words, the projection of the nonlinear source term on the modal basis is not fully along the fundamental mode but it is rather composed of a lot of non zero components on the higher order modes. Accordingly, higher order modes are created by the nonlinear propagation. In the present propagation example, the second transversal mode is the most efficiently populated, explaining in turn the periodicity of the fluence distribution. More particularly, the transversal radius of the pulse decreases and then recovers its initial size periodically, which is directly linked to the fact that the second mode population periodically increases and decreases due to the phase mismatch along the propagation. This phenomenon is the same as the one explaining why temporal function is shorter and shorter if its frequency spectrum is larger and larger: the transversal size of a field propagating in a waveguide is smaller and smaller if it is composed of more and more in-phase transversal modes, the latter playing the role of the transversal spectrum.
\begin{figure}
	\includegraphics[width=9cm,keepaspectratio]{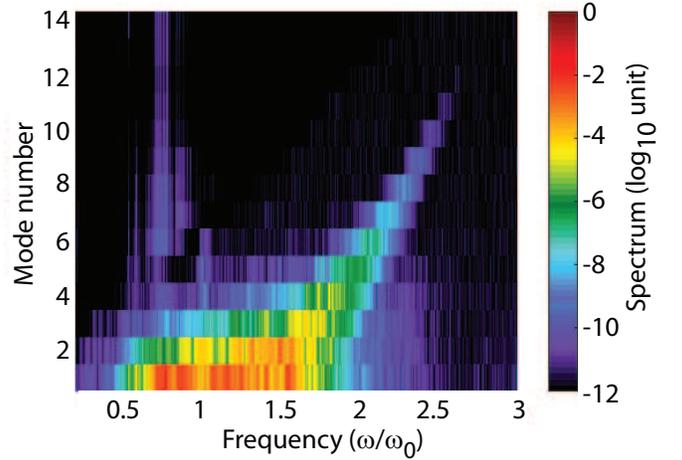}
	\caption{Mode-resolved spectrum after a 10\,cm propagation in a 20\,$\mu$m step-index silica fiber.}
	\label{Fig9}
\end{figure}
In the above discussion, both the generation of the supercontinuum and the evolution of the transversal distribution along the propagation have been described in a complete uncorrelated manner. However, if one now looks at how the pulse spectrum is distributed along the transversal modes (Fig.\,\ref{Fig9}) after propagation, it appears that strong spatio-temporal couplings take place during the propagation. The modal spectrum indeed exhibits a structure similar to the fish waves generated close to the zero-dispersion wavelength in bulk materials. The redder part of generated supercontinuum mainly lies in the fundamental transversal mode while the bluer part is mainly distributed in the higher order modes. While the theoretical analysis and discussion about spatially and temporally localized conical waves in nonlinear guided media will be the subject of another publication, it is nevertheless particularly interesting to underline the parallel between the propagation in bulk media and in multimode fibers.
Note also that preliminary studies also show that X-waves (resp. O-waves) are generated in the normal (resp. anomalous) group velocity dispersion regime (data not shown), making the parallel even more obvious.
\section{Conclusion}
In conclusion, we have derived the general unidirectional pulse propagation equation driving the evolution of an electric field propagating in a waveguide, which generalizes the UPPE obtains in bulk materials. Moreover, we have presented an original method for both solving the propagation equation but also for evaluating the optical modes supported by the waveguide for any kind of geometry. In particular, a fast modal transform based on the fast Hankel transform has been presented. Finally, as an example, we have solved the propagation equation in a multimode step index fiber and have demonstrated that conical waves can be formed during the propagation, as it is the case in bulk media.
\acknowledgments
This work was supported by the Conseil R\'egional de Bourgogne (PARI program) and the CNRS. We thank the CRI-CCUB for CPU loan on the multiprocessor server.

\bibliographystyle{unsrt}

\begin{thebibliography}{1}
\bibitem{1} L.G. Wright, D.N. Christodoulides, and F.W. Wise, Controllable spatiotemporal nonlinear effects in multimode fibres. Nat. Phot. \textbf{9}, 306-310 (2015).
\bibitem{2} A. Picozzi, G. Millot, and S. Wabnitz, Nonlinear virtues of multimode fibre. Nat. Phot. \textbf{9}, 289–291 (2015).
\bibitem{3} L.G. Wright, Z. Liu, D.A. Nolan, M.-J. Li, D. N. Christodoulides, and F.W. Wise, Self-organized instability in graded-index multimode fibres. Nat. Phot. \textbf{10}, 771-776 (2016)
\bibitem{4} W.H. Renninger and F.W. Wise, Optical solitons in graded-index multimode fibres. Nat. Comm. \textbf{4}, 1719 (2013).
\bibitem{5} K. Krupa, A. Tonello, B.M. Shalaby, M. Fabert, A. Barthelemy, G. Millot, S. Wabnitz, and V. Couderc, Spatial beam self-cleaning in multimode fibres. Nat. Phot. \textbf{11}, 237-241 (2017).
\bibitem{6} P. St. J. Russell, P. Holzer, W. Chang, A. Abdolvand, and J. C. Travers, Hollow-core photonic crystal fibres for gas-based nonlinear optics. Nat. Phot. \textbf{8}, 278-286 (2014).
\bibitem{7} P. Bejot, B. E. Schmidt, J. Kasparian, J.-P. Wolf, and F. Legare, Mechanism of hollow-core-fiber infrared-supercontinuum compression with bulk material. Phys. Rev. A \textbf{81}, 063828 (2010).
\bibitem{8} X. Chen, A. Malvache, A. Ricci, A. Jullien, and R. Lopez-Martens, Efficient hollow fiber compression scheme for generating multi-mJ, carrier-envelope phase stable, sub-5 fs pulses. Laser Physics \textbf{21}(1), 198–201 (2011).
\bibitem{9} T.G. Philbin, C. Kuklewicz, S. Robertson1, S. Hill, F. Konig, U. Leonhardt, Fiber-Optical Analog of the Event Horizon. Science \textbf{319}(5868), 1367-1370 (2008).
\bibitem{10} J. Andreasen and M. Kolesik, Nonlinear propagation of light in structured media: Generalized unidirectional
pulse propagation equations. Phys. Rev. E \textbf{86}, 036706 (2012).
\bibitem{11} F. Poletti and P. Horak, Description of ultrashort pulse propagation in multimode optical fibers. JOSA B \textbf{25}(10), 1645-1654 (2008).
\bibitem{12} M. Guizar-Sicairos and J. C. Gutierrez-Vega, Computation of quasi-discrete Hankel transforms of integer order for propagating optical wave fields. JOSA A \textbf{21}(1) 53-58 (2004).
\bibitem{13} D. Faccio, A. Averchi, A. Couairon, M. Kolesik, J.V. Moloney, A. Dubietis, G. Tamosauskas, P. Polesana, A. Piskarskas, and P.Di
Trapani, Spatio-temporal reshaping and X Wave dynamics in optical filaments. Opt. Express \textbf{15}(20) 13077-13095 (2007).
\bibitem{14} C. Conti, Generation and nonlinear dynamics of X waves of the Schrödinger equation. Phys. Rev. E \textbf{70}, 046613 (2004)
\bibitem{15} P. Di Trapani, G. Valiulis, A. Piskarskas, O. Jedrkiewicz, J. Trull, C. Conti, and S. Trillo, Spontaneously Generated X-Shaped Light Bullets. Phys. Rev. Lett. \textbf{91} 093904 (2003).
\bibitem{16} E. Arevalo, Boosted X Waves in Nonlinear Optical Systems. Phys. Rev. Lett. \textbf{104}, 023902 (2010).
\bibitem{17} L. G. Wright, S. Wabnitz, D. N. Christodoulides, and F. W. Wise, Ultrabroadband Dispersive Radiation by Spatiotemporal Oscillation of Multimode Waves. Phys. Rev. Lett. \textbf{115}, 223902 (2015).
    \bibitem{18} D. Hollenbeck and C. D. Cantrell, Multiple-vibrational-mode model for fiber-optic Raman gain spectrum and response function. JOSA B \textbf{19}(12) 2886-2892 (2002).
\end{thebibliography}

\end{document}